\documentclass[aps,twocolumn,showpacs,showkeys]{revtex4}
\usepackage{graphicx}
\usepackage{latexsym}
\newlength{\oldunit}
\begin{document}
\title{Irreversible Quantum Baker Map}
\author{Artur {\L}ozi\'nski} \email{lozinski@if.uj.edu.pl}
\author{Prot Pako\'nski} \email{pakonski@if.uj.edu.pl}
\affiliation{Uniwersytet Jagiello\'nski,
     Instytut Fizyki im. M.~Smoluchowskiego, \\
     ul. Reymonta 4, 30--059 Krak\'ow, Poland}
\author{Karol \.Zyczkowski} \email{karol@cft.edu.pl}
\affiliation{Centrum Fizyki Teoretycznej, Polska Akademia Nauk, \\
     Al. Lotnik\'ow 32/44, 02--668 Warszawa, Poland}
\begin{abstract}
We propose a generalization of the model of classical baker map
on the torus, in which the images of two parts of the phase space
do overlap.  This transformation is irreversible and cannot be
quantized by means of a unitary Floquet operator.  A
corresponding quantum system is constructed as a completely
positive map acting in the spa\-ce of density matrices.  We
investigate spectral properties of this super-operator and their
link with the increase of the entropy of initially pure states.
\end{abstract}
\maketitle

In last years a re-exploration of a finite dimensional Hilbert
space ${\cal H}$ and the space of density operators acting on it
took place.  Emerging interest in properties of mixed quantum
states is stimulated by research on the decoherence
phenomena~\cite{Zurek,Haroche} and recent developments in modern
applications of quantum mechanics including quantum information,
cryptography and computing~\cite{Nielsen}.  The concept of  mixed
states is crucial while analyzing non-unitary quantum evolution,
necessary to describe processes of measurement and interaction
with an environment~\cite{Bennett,Vincenzo}.

Any quantum map ${\mathbf{ \Lambda}}$ should send positive
density o\-pe\-ra\-tors into other positive operators. Moreover, since
any system under consideration, described by a density operator,
may be coupled to an environment, so ${\mathbf{ \Lambda \otimes
1_m}}$ should be positive for any extension of ${\mathbf{
\Lambda}}$ by the m-dimensional identity matrix ${\mathbf{
1_m}}$. This property is called {\sl complete positiveness}
(CP)~\cite{Kraus}. If the classical dynamics preserves
probability, then the corresponding
quantum map should preserve the trace of the density operator.

Research on quantum analogs of classically chaotic dynamical
system often concentrates on 2-d area preserving maps. The most
popular examples include the classical baker map and the Arnold cat
map. They were quantized by finding the corresponding unitary operators,
which act on a finite dimensional Hilbert space ${\cal H}$
(see e.g. \cite{SS00} and references therein).

In this letter we propose a generalization of the classical and
quantum baker map. The classical map proposed is irreversible,
and therefore its quantum counterpart cannot be represented by a
unitary operator.  The classical map transforms a unit square
into a rectangular subset of it, while the quantum map is a
completely positive, trace preserving super-operator acting in
the space of density matrices of a fixed size. Our research is
related to recent papers of Soklakov and Schack \cite{SS00}, and
Saraceno \emph{et.al.} who quantized a dissipative version of the
baker map~\cite{Vallejos} and also studied a stochastic system
devised to take into account the effects of
decoherence~\cite{Bianucci}. However, the system analyzed here is
different, since it is not dissipative, it conserves the
probability and is deterministic. To demonstrate that the quantum
dynamics corresponds to the classical one, we compare the quantum
and the classical evolution of an initial density and show
quantum counterparts of the classical periodic orbits.

\begin{figure}[hbt]
  \setlength{\oldunit}{\unitlength}
  \setlength{\unitlength}{1.3ex}
  \begin{center}
    \begin{picture}(45,10)
\put(4.5,-0.7){\mbox{\large $q$}}
\put(33,1){\begin{picture}(8,8)
\put(0,6){\line(0,1){2}}
\put(0,8){\line(1,0){8}}
\put(8,6){\line(0,1){2}}
\put(0,0){\begin{picture}(8,4)
\put(0,0){\line(1,0){8}}
\put(0,0){\line(0,1){4}}
\put(0,4){\line(1,0){8}}
\put(8,0){\line(0,1){4}}
\put(0,3){\line(4,1){4}}
\put(0,2){\line(4,1){8}}
\put(0,1){\line(4,1){8}}
\put(0,0){\line(4,1){8}}
\put(4,0){\line(4,1){4}}
\end{picture}}
\put(0,2.5){\begin{picture}(8,4)
\put(0,0){\line(1,0){8}}
\put(0,0){\line(0,1){4}}
\put(0,4){\line(1,0){8}}
\put(8,0){\line(0,1){4}}
\put(0,1){\line(4,-1){4}}
\put(0,2){\line(4,-1){8}}
\put(0,3){\line(4,-1){8}}
\put(0,4){\line(4,-1){8}}
\put(4,4){\line(4,-1){4}}
\end{picture}}
\put(8,-1.5){\mbox{1}}
\put(-1,-1.5){\mbox{0}}
\put(-1,8){\mbox{1}}
\put(10,2.5){\line(0,1){1.5}}
\put(9.6,4){\line(1,0){0.8}}
\put(9.6,2.5){\line(1,0){0.8}}
\put(11,2.55){$\Delta /2$}
\end{picture}}
\put(26,5){\vector(1,0){6}}
\put(28.1,5.6){$L_\Delta$}
\put(17,1){\begin{picture}(8,8)
\put(0,6){\line(0,1){2}}
\put(0,8){\line(1,0){8}}
\put(8,6){\line(0,1){2}}
\put(0,0){\begin{picture}(8,4)
\put(0,0){\line(1,0){8}}
\put(0,0){\line(0,1){4}}
\put(0,4){\line(1,0){8}}
\put(8,0){\line(0,1){4}}
\put(0,3){\line(4,1){4}}
\put(0,2){\line(4,1){8}}
\put(0,1){\line(4,1){8}}
\put(0,0){\line(4,1){8}}
\put(4,0){\line(4,1){4}}
\end{picture}}
\put(0,4){\begin{picture}(8,4)
\put(0,0){\line(1,0){8}}
\put(0,0){\line(0,1){4}}
\put(0,4){\line(1,0){8}}
\put(8,0){\line(0,1){4}}
\put(0,1){\line(4,-1){4}}
\put(0,2){\line(4,-1){8}}
\put(0,3){\line(4,-1){8}}
\put(0,4){\line(4,-1){8}}
\put(4,4){\line(4,-1){4}}
\end{picture}}
\put(8,-1.5){\mbox{1}}
\put(-1,-1.5){\mbox{0}}
\put(-1,8){\mbox{1}}
\end{picture}}
\put(10,5){\vector(1,0){6}}
\put(12.1,5.6){$\Theta$}
\put(-0.9,4.5){\mbox{\large $p$}}
\put(1,1){\begin{picture}(8,8)
\put(0,0){\begin{picture}(4,8)
\put(0,0){\line(1,0){4}}
\put(0,0){\line(0,1){8}}
\put(0,8){\line(1,0){4}}
\put(4,0){\line(0,1){8}}
\put(2,0){\line(1,1){2}}
\put(0,0){\line(1,1){4}}
\put(0,2){\line(1,1){4}}
\put(0,4){\line(1,1){4}}
\put(0,6){\line(1,1){2}}
\end{picture}}
\put(4,0){\begin{picture}(4,8)
\put(0,0){\line(1,0){4}}
\put(0,0){\line(0,1){8}}
\put(0,8){\line(1,0){4}}
\put(4,0){\line(0,1){8}}
\put(2,0){\line(-1,1){2}}
\put(4,0){\line(-1,1){4}}
\put(4,2){\line(-1,1){4}}
\put(4,4){\line(-1,1){4}}
\put(4,6){\line(-1,1){2}}
\end{picture}}
\put(8,-1.5){\mbox{1}}
\put(-1,-1.5){\mbox{0}}
\put(-1,8){\mbox{1}}
\end{picture}}
   \end{picture}
   \caption{Classical sloppy baker map: after the original baker
            transformation $\Theta$, the top half of the square is
            shifted down by $\frac{\Delta}{2}$ (operator
     $L_\Delta$).}
    \label{pic_tr}
  \end{center}
\end{figure}
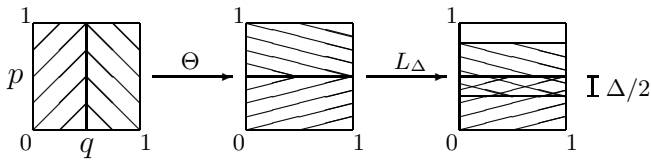

The standard baker map is a transformation of the unit square $I$,
a model of a finite phase space, onto itself. It
consists of stretching the square in one direction, labeled by $q$,
and squeezing it in another direction (labeled by $p$) by the factor of $2$.
After the stretching procedure the
baker cuts the rectangle into two pieces and places the right piece at
the top of the left one, as shown in Fig.~\ref{pic_tr} (transformation
$\Theta$). Assume that instead of doing this, the sloppy baker puts the
right piece a bit too low, in such a way that a $\frac{\Delta}{2}$
overlap with the left piece occurs. This effect is described by the
transformation $L_\Delta$ (formally an interval translation map
acting in the $p$ direction~\cite{Boshernitzan}), which shifts all points
from the top half of the square\\ ($p>1/2$) down by
$\frac{\Delta}{2}$. The formal definition of the \emph{classical
sloppy baker map} is
\begin{equation} \Theta_\Delta :
\left( \begin{array}{l}
q\\
p
\end{array}\right)
\stackrel{B}{\rightarrow}
\left( \begin{array}{l}
2q-[2q]\\
\frac{1}{2}(p+[2q])
\end{array}\right)
\stackrel{L_\Delta}{\rightarrow}
\left( \begin{array}{l}
2q-[2q]\\
\frac{1}{2}(p+[2q](1-\Delta))
\end{array}\right) ,
\label{theta_d}
\end{equation}
where $[x]$ denotes the integer part of $x$ and the parameter
$\Delta$ belongs to  $[0,1]$.  The map $\Theta_\Delta$ is not
reversible for $\Delta>0$, because any point for which $p\in
(\frac{1-\Delta}{2},\frac{1}{2}]$ has two preimages, while the
points with $p> 1-\frac{\Delta}{2}$ have none.

We will use density distribution $f$ on the square, $\int _I f(\alpha
)\, \mathrm{d}\alpha \, = \, 1,$ $f \geq 0$, where $\alpha$ is a short
notation for the pair $(q,p)$. The map $\Theta_\Delta$ generates the
Frobenius-Perron operator acting in the space of classical density
distributions
\begin{equation} \label{fro-per}
{\cal M}f(\alpha ) =
\int _I f(\alpha ^{'} )\delta (\alpha - \Theta(\alpha ^{'} ))
\, \mathrm{d}\alpha ^{'} .
\end{equation}
Since the map is not dissipative, and $\Theta_\Delta (I) \subset I $, the
operator ${\mathcal{M}}$ preserves the probability, $\int_I
{\mathcal{M}}f(\alpha)\, \mathrm{d}\alpha = \int_I f(\alpha )\,
\mathrm{d}\alpha = 1$.
The density $f^\ast (\alpha ) = \frac{1}{1-\Delta}$ for
\mbox{$\alpha \in [0,1]\times [0,1-\Delta)$} and $0$ elsewhere is invariant under
the action of ${\mathcal{M}}$ and ${\mathcal{M}}f^\ast = f^\ast$.
Several versions of quantum baker map on the torus are
known~\cite{Balazs,Saraceno,POZ00}.  We
use the first form of the quantum operator proposed by Balazs and
Voros~\cite{Balazs}
\begin{equation}
{\mathbf{B}} = F ^{\dagger}_{N} \cdot \left( \begin{array}{cc}
F_{N/2} & 0 \\
0 & F_{N/2}
 \end{array} \right) ,
\end{equation}
where $F_{N}$ denotes the N-points discrete Fourier transformation,
$[F_N]_{kl} = \frac{1}{\sqrt{N}}\, e^{-2\pi i kl/N}$.
Since the sloppy map $\Theta_\Delta$ does not enjoy the symmetry
of the original baker map, we will not need the symmetric quantum model
introduced by Saraceno~\cite{Saraceno}. Unitary operator
${\mathbf{B}}$ acts on the $N$ dimensional Hilbert space ${\cal H}_N$,
where $N$ is even.
\begin{figure}[hbt]
  \begin{center}
    \includegraphics[width=.325\textwidth]{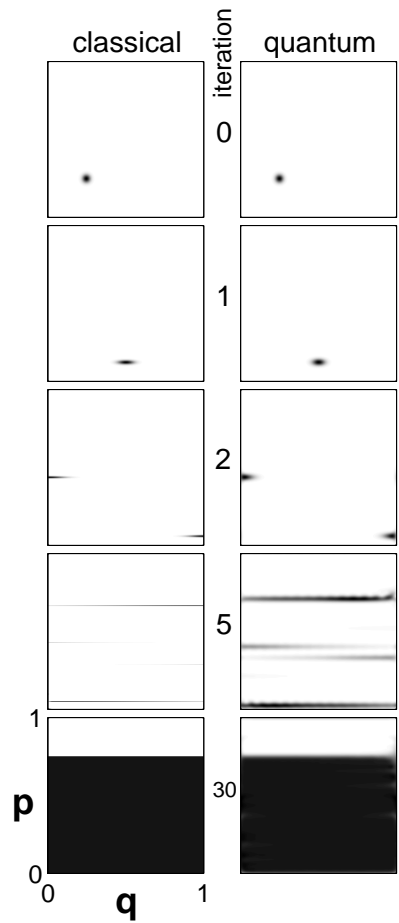}
    \caption{Sloppy baker map with $\Delta=1/4$.
     Time evolution of an initially localized classical density
     concentrated at $\alpha_0=(0.25,0.25)$ (left) and Husimi representation
	  (\ref{husro}) of the density
     matrix of an initially pure state localized in the same
     point $\alpha_0$ and iterated by the quantum map (\ref{Bdelta})
     for $N=512$  (right).}
    \label{pic_ev}
  \end{center}
\end{figure}

The classical map $\Theta_\Delta$ is irreversible, so its
quantization cannot be achieved by means of unitary operators.
The quantum operator ${\mathbf{\Lambda_\Delta}}$ ---
corresponding to the classical map $L_\Delta$ should act on the
space of mixed quantum states and may be realized by a
super-operator. Any super-operator (${\mathbf{\Lambda}}$) which
defines a completely positive map, may be written in the
so-called Kraus form~\cite{Kraus}
\begin{equation} \label{superoperator}
  {\mathbf{\Lambda}} (\rho ) = \sum _{i=1}^{K} A_i\rho A^{\dagger}_i ,
\end{equation}
where $\rho$ is a density matrix and $K$ is finite.
If operators $A_i$ fulfill the
condition
\begin{equation} \label{consistency}
  \sum _{i=1}^{K} A^{\dagger}_i A_i = {\mathbf{1}}_N,
\end{equation}
the map ${\mathbf{\Lambda}}(\rho)$ is trace preserving.  The
classical map $\Theta_\Delta$ transforms the bottom half of the
square $I$ into itself and shifts the top one down by
$\frac{\Delta}{2}$. The two halves of $I$ are transformed
separately.  Therefore we split the phase space into bottom and
top and introduce two projection operators $D_b$ and $D_t$,
which written in the eigenbasis of position operator have the
form
\begin{equation}
  D_b = F^{\dagger}_N \left( \begin{array}{cc}
	{\mathbf{1}}_{N/2}&0\\
	0&0
  \end{array} \right) F_N; \quad
  D_t = F^{\dagger}_N \left(
  \begin{array}{cc}
	0&0\\
	0&{\mathbf{1}}_{N/2}
  \end{array} \right) F_N,
\end{equation}
Notice that the super-operator ${\mathbf{\Lambda_M}}(\rho) =
D_b\rho D^{\dagger}_b + D_t\rho D^{\dagger}_t $ corresponds to the up/down
measurement process and the Kraus operators $A_1 = D_b$ and $A_2 = D_t$
fulfill the condition (\ref{consistency}). To construct a quantum shift
transformation ${\mathbf{\Lambda_\Delta}}$ we will use the unitary
operator of translation in momentum,
\begin{equation}
  V | k \rangle = | k+1 \rangle , \quad V^N = {\mathbf{1}}_N.
\label{ptrans}
\end{equation}
Here $| k \rangle$ denotes the discrete eigenstate of momentum
which is periodic, $| k+N \rangle = | k \rangle$~\cite{Saraceno}.
For $k=1$ the state is localized at the bottom of $I$.
Then the  vertical shift of the the top half of $I$ by $\Delta/2$
is realized by the translation operator (\ref{ptrans}) acting on
the previously measured system,
\begin{equation}
  D'_t = V^{-N\Delta/2} \cdot D_t .
\end{equation}
For simplicity we assure here that exponent is integer, but the
same construction works also for any real $\Delta$.
Since the position of the bottom part remains unchanged,
the entire quantum
transformation ${\mathbf{\Lambda_\Delta}}$ reads
\begin{equation}
  {\mathbf{\Lambda_\Delta}}(\rho) = D_b \rho D_b^\dagger
+ D'_t \rho D'^\dagger_t  .
\end{equation}
This super-operator resets to 0 the off-diagonal blocks of the
$\rho$ matrix in the $p$-representation. This is related to the
fact that to displace one half of the torus we need to perform a
measurement, which implies decoherence.  Thus even
for $\Delta=0$ the operator ${\mathbf{\Lambda_\Delta}}$ differs
from identity, but the effect of the measurement becomes
negligible in the classical limit $N\to \infty$.

Using the above super-operator (${\mathbf{\Lambda_\Delta}}$) we
construct the entire quantum sloppy baker map
\begin{equation}
  {\mathbf{B_\Delta}}(\rho) = {\mathbf{\Lambda_\Delta}}(B \rho B^\dagger)
    = D_b B \rho B^\dagger D_b^\dagger
    + D'_t B \rho B^\dagger D'^\dagger_t \ .
\label{Bdelta}
\end{equation}
Note that the Kraus operators $A_1=D_bB$ and $A_2=D'_tB$
fulfill condition (\ref{consistency}) with $K=2$.

To demonstrate that quantum system defined by (\ref{Bdelta})
corresponds to the classical sloppy baker map we compare
classical and quantum structures in the phase space. In order to
define quantum quasi-probability distribution, we
use a family of states localized at points of the square $N
\times N$ lattice in the phase space constructed by means of
translation operators~\cite{Saraceno}.  The operator $U$ of
translation in position is defined similarly to $V$,
\begin{equation}
  U | n \rangle = | n+1 \rangle , \quad U^N = {\mathbf{1}}_N,
\end{equation}
where $|n\rangle$ are position eigenstates, satisfying
$|n+N\rangle = |n\rangle$. As a reference state we choose
arbitrarily the wave packet $|\frac{1}{2},\frac{1}{2}\rangle$
localized in $(\frac{1}{2},\frac{1}{2})$
\begin{equation}\label{gstate}
  \langle n | 1/2,1/2 \rangle = (2/N)^{-1/4}e^{-\pi (n-N/2)^2/N-i\pi n}
\end{equation}
which becomes Gaussian for $N\rightarrow \infty$.
We translate it to any point $(q,p)$, where $Nq$ and $Np$ are
integers ($N$ is even)
\begin{equation}
  | q,p \rangle = V^{Np-N/2} \; U^{Nq-N/2} \; | 1/2,1/2 \rangle .
\end{equation}
These states allow one to define a Husimi
representation in the phase space of any mixed quantum state
$\rho$
\begin{equation} \label{husro}
  H_{\rho}(q,p ) = \langle q,p| \; \rho \; |q,p\rangle .
\end{equation}

We analyzed the evolution of an exemplary state
$|\alpha_0\rangle$ localized at $\alpha_0$ and the classical
transformation of the corresponding density distribution. On the
left hand side of Fig.~\ref{pic_ev} we present the classical
density and its image after $T=1,2,5$ and $T=30$ iterations of
the Frobenius-Perron operator (\ref{fro-per}). The right hand
side shows the Husimi representations (\ref{husro}) of the
initially pure quantum state $|\alpha_0\rangle\langle\alpha_0|$
and its images after $T$ actions of the super-operator
${\mathbf{B_\Delta}}$.  The quantum quasi-probability
distribution $H_\rho$ is localized in the same region of the
phase space as the classical density distribution. Since the
Husimi distribution may resolve quantum phase-space structures at
the length scale of the order of $\hbar^{1/2}\propto N^{-1/2}$,
the classical density becomes narrower then its quantum
counterpart already after first iteration.
\begin{figure}[hbt]
  \begin{center}
    \includegraphics[width=.325\textwidth]{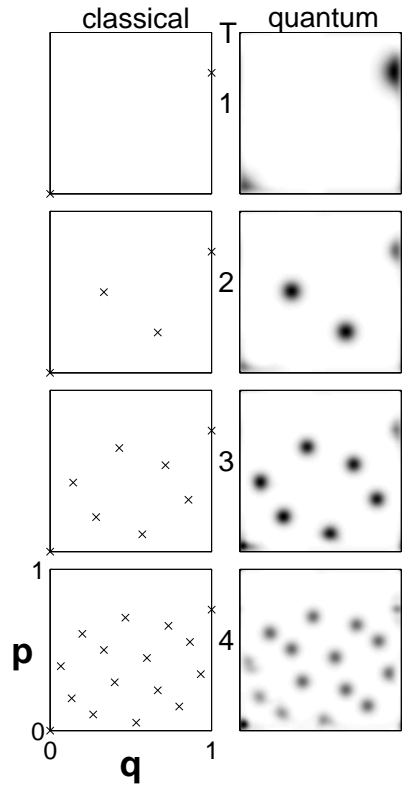}
    \caption{Classical orbits of period $T=1,2,3,4$ ($\times$)
      (left) are localized close to the peaks of the quantum
      return probability $R^T$ (right) obtained for the sloppy
      baker map with $\Delta=1/4$ and $N=96$.}
    \label{pic_orb}
  \end{center}
\end{figure}

After 30 iterations of the classical map the density
distribution is close to the invariant measure $f^\ast$. Also the
quantum state ${\mathbf{B_\Delta}}^{30}(|\alpha_0\rangle
\langle\alpha_0|)$ is close to the invariant density matrix
\begin{equation} \label{invard}
  \rho^\ast = {\mathbf{B_\Delta}} (\rho^\ast),
\end{equation}
the existence of which is guaranteed by the trace preserving
condition (\ref{consistency}). The state $\rho^\ast$ is localized
on the rectangle $[0,1]\times[0,1-\Delta]$. Moreover, it is
almost isotropic on the corresponding $N(1-\Delta)$ dimensional
subspace. To show this we verified that the von Neumann entropy
of the invariant state $\mbox{S}(\rho^\ast)=-\mbox{Tr}\,
\rho^\ast\ln\rho^\ast$, is close to the maximal entropy for the
$N(1-\Delta)$ dimensional subspace of the Hilbert space
${\mathcal{H}}_N$,
\begin{equation}
  \mbox{S}(\rho^\ast) \approx \mbox{S}_{\rm max}^{[N(1-\Delta)]}:=\ln(N(1-\Delta)).
    \label{inv_ent}
\end{equation}

It is instructive to look at the periodic orbits of the classical
transformation $\Theta{}_\Delta$. They are those of the original
(reversible) baker map with momentum scaled by the factor
$(1-\Delta)$,
\begin{equation}
  q_T^{\ast} = \frac{n}{2^T-1} \ , \quad
  p_T^{\ast} = \frac{r(n)}{2^T-1} (1-\Delta) \ ,
\end{equation}
where $T$ denotes the length of the period, $n$ ranges from $0$
to $2^T-1$, and the symbol $r(n)$ denotes the number obtained
from $n$ by reversing the order of its bits.  The classical
periodic orbits  may be compared with the structures of the
quantum return probability
\begin{equation} \label{return}
  R^T(q,p) = \langle q,p | {\mathbf{B_\Delta}}^T(|q,p\rangle\langle q,p|)
     |q,p \rangle .
\end{equation}
The function $R^T(q,p)$ measures the projection of the quantum
state $|q,p\rangle\langle q,p|$ iterated $T$ times by the
super-operator ${\mathbf{B_\Delta}}$ onto itself. As shown in
Fig.~\ref{pic_orb} its maxima are indeed located in the
vicinity of classical periodic orbits.

\begin{figure}
  \begin{center}
    \includegraphics[width=.48\textwidth]{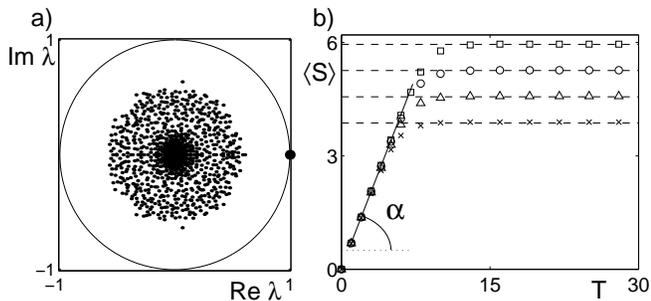}
    \caption{a) Eigenvalues of the super-operator
      ${\mathbf{B_\Delta}}$ for $N=64$ and $\Delta=1/4$ in the
      complex plane, larger dot denotes $\lambda _1$.
		b) Dependence of the mean von Neumann
      entropy $\langle \mbox{S}\rangle$ on time $T$ for the
		irreversible quantum baker
      map with $\Delta=1/4$ and $N=64 (\times), 128 (\triangle),
      256 (\circ)$ and $512 (\Box)$. Horizontal lines represent
      asymptotic estimation (\ref{inv_ent}).}
    \label{pic_ent}
  \end{center}
\end{figure}

Spectral decomposition of the superoperator ${\mathbf{B_\Delta}}$
determines the time evolution of the system. For any trace
preserving CP map~(\ref{superoperator}) the operator
${\mathbf{\Lambda}}$ has an eigenvalue $\lambda_1=1$
corresponding to the invariant state $\rho^\ast$.  The spectrum
is symmetric with respect to the real axis, since
${\mathbf{\Lambda}}$ \mbox{sends} Hermitian density matrices into
density matrices \cite{TDV00}.  Not every superoperator needs to
be diagonalizable, i.e.  the number of eigenvectors may be
smaller than the number of eigenvalues.  This is the case for the
super--operator of translation  ${\mathbf{\Lambda_\Delta}}$, the
spectrum of which consists of two eigenvalues $0$ and $1$. The
multiplicity of the former is equal to $3N^2/4$ and the
corresponding subspace is defective for any $\Delta>0$.

Fig.~\ref{pic_ent}a shows all $N^2$ eigenvalues of the linear
operator ${\mathbf{B_\Delta}}$ for $N=64$ and $\Delta=1/4$.
Observe a considerable {\sl spectral gap}, i.e. the difference
$1-|\lambda_2|$, which determines the rate of the convergence of
an initial state toward the invariant density matrix
$\rho^\ast$. Moduli of the largest subleading eigenvalues
influence the slope $\alpha$ of the initially linear
entropy increase demonstrated in  Fig.~\ref{pic_ent}b.  The data
were obtained by averaging over a sample of $10$ initially
pure states drawn  randomly with respect to the unique, unitarily
invariant measure on the $2N-2$ dimensional space of pure states in
${\mathcal{H}}_N$.

In this work we introduced an irreversible baker map and proposed
a method of its quantization. On one hand, the limit
$N\rightarrow\infty$ of the model may
be useful to analyze the quantum--classical correspondence for
chaotic, completely positive quantum maps. On the other hand, the
extreme quantum regime of low $N$ may be interesting from the
point of view of quantum information. Quantum baker map becomes a
standard model for  theoretical \cite{TS99} and experimental
\cite{WLEC02} investigation of NMR quantum computing, and our
generalization makes it possible to study the consequences of
irreversibility in the system.  The effects of the decoherence
and the dynamics of entanglement in the two--qubits version of
this system ($N=4$) as well as a generalization of the model will
be presented elsewhere.  It is a pleasure to thank R.~Alicki,
A.~Becker, M.~Ku{\'s}, F.~Minert, R.~Rudnicki and D.~W{\'o}jcik
for fruitful discussions. This work was supported by Polish KBN
grant no 2P03B-072-19.

\end{document}